\documentclass[amsmath,amssymb,pra,twocolumn]{revtex4}
\usepackage[cp1251]{inputenc}
\usepackage[english]{babel}
\usepackage{color}
\usepackage{graphicx}
\usepackage{dcolumn}
\usepackage{braket}
\usepackage{amsmath}

\usepackage{natbib}

\begin{document}
\bibliographystyle{unsrt}

\title{Entanglement of multiphoton two-mode polarization Fock states and of their superpositions.}

\author{S. V. Vintskevich$^{1,3}$, D. A. Grigoriev$^{1,3}$, N.I. Miklin$^4$ , M. V. Fedorov$^{1,2}$ }
\address{$^1$A.M.~Prokhorov General Physics Institute,
Russian Academy of Sciences, 38 Vavilov st., Moscow, 119991, Russia}
\address{$^2$National Research University Higher School of Economics, 20
Myasnitskaya Ulitsa,
Moscow, 101000, Russia}
\address{$^3$Moscow Institute of Physics and Technology, Dolgoprudny, Moscow Region, Russia}
\address{$^4$ Institute of Theoretical Physics and Astrophysics, National Quantum Information Center, Faculty of Mathematics, Physics and Informatics, 80-308, Gdansk, Poland}

\date{\today}

\begin{abstract}
Density matrices of pure multiphoton Fock polarization states and of arising from them reduced density matrices of mixed states are expressed in similar ways in terms of matrices of correlators defined as averaged products of equal numbers of creation and annihilation operators. Degree of entanglement of considered states is evaluated for various combinations of parameters of states and character of their reduction.

\end{abstract}

\maketitle

\section{Introduction}
There is a growing interest in the science of quantum information to multiparticle entanglement. It finds applications in quantum computing and quantum error correction \cite{Shor},\cite{Briegel}, as well as in quantum networks \cite{Kimble}. The latter includes, in particular, communication among many parties that is enhanced by shared multiparticle entanglement. The most promising recourse for establishing this type of entanglement are, of course, multi-photon systems. Thus, there is a natural interest in studying entanglement properties of states of many photons. Recent experiments also shown that entanglement of up to ten photons can be observed in a lab \cite{Wang}.

In this work we consider pure multiphoton two-mode polarization Fock states and their superpositions. We give a general definition of the density matrices of such states as well as of the density matrices of mixed states arising from pure Fock states after their partial reduction over a series of photon variables. Elements of such density matrices are expressed in terms of correlators defined as averaged products of equal numbers of creation and annihilation operators with different distributions of operators over two polarization modes. we will calculate parameters characterizing the degree of entanglement in such states and investigate their dependence on features of the original pure states and on the ways of their reduction.

Note that for biphoton states the method of density matrices of the described type was suggested
by D.N. Klyshko in 1997 \cite{Klyshko} and somewhat later used in the works \cite{Che-Burl,Che-Kul}. More recently there was a series of works on some aspects of entanglement in  multipohoton states \cite{Wu,Dalton,KWang,Lu,Mikl}. But as far as we know, there were no works where the Klyshko method of density matrices would be generalized for multiphoton states with numbers of photons higher than 3. Such generalization is one of the main goals of this work. The second goal is characterizing entanglement of multiphoton states in terms of Schmidt decompositions and their parameters, which will be new too. Note also that, though the Schmidt decomposition was known in mathematics since 1906 \cite{Schm}, in the fields of modern quantum optics and quantum information it was introduced by J.H. Eberly and coworkers at first in 1994 \cite{grobe} and then in 2004 \cite{LE}. A much more general and detailed description of the Schmidt decomposition, as well as its applications, were given in the review paper \cite{CP}.

\section{Density matrices}
Let us consider an arbitrary pure state $\ket{\Psi^{(n)}}$ of $n$ photons having identical frequencies and identical given propagation directions but distributed arbitrarily between two polarization modes, horizontal and vertical ones, $H$ and $V$. Two-mode polarization basic Fock states are states with given numbers of horizontally and vertically polarized photons $n_H$ and $n_V$ such that $n_H+n_V=n$
\begin{equation}
 \label{basis}
 \ket{\Psi_{n_H,\,n_V}}=\ket{n_H,n_V}
 =\frac{a_H^{\dag\,n_H}a_V^{\dag\,n_V}}{\sqrt{n_H!n_V!}}\ket{0}.
\end{equation}
More general $n$-photon polarization states to be considered are superpositions of basic Fock states
\begin{equation}
 \label{Sup}
 \ket{\Psi^{(n)}}=\sum_{n_H=0}^n  C_{n_H}\left.\ket{n_H,n_V}\right|_{n_V=n-n_H}
\end{equation}
with $\sum_{n_H=0}^n |C_{n_H}|^2=1$.
The wave functions of all $n$-photon states $\ket{\Psi^{(n)}}$ depend on $n$ single-photon variables $\sigma_i$, $\Psi^{(n)}(\{\sigma_i\})=\braket{\{\sigma_i\}|{\Psi^{(n)}}}$ and, explicitly, they are given by symmetrized products of $n$ single-photon wave functions \cite{Schweber}. In the case of polarization modes the single-photon wave functions in these products are $\psi_H(\sigma_i)=\delta_{\sigma_i,\,H}$ and $\psi_V(\sigma_j)=\delta_{\sigma_j,\,V}$. In the matrix representation $\psi_H(\sigma_i)=\left(1\atop 0\right)_i$ and $\psi_V(\sigma_j)=\left(0\atop 1\right)_j\,$ \cite{JETP}.

Note that sometimes it's possible to meet in literature mentions about particle- or mode- entanglement and about differences or  similarities between them. We do not use such concepts here because in our opinion the type of entanglement to be studied can be much more correctly interpreted as related to uncertainty of distributions of particle variables between modes or, shortly, as the variable entanglement. For two-mode polarization states this means an uncertainty of attachment of polarization variables $\sigma_i$ to $H$- or $V$-modes.

Direct products of $n$ two-line columns $\left(1\atop 0\right)_i$ and $\left(0\atop 1\right)_j\,$ form a basis of columns with $2^n$ elements (``rows$"$) and with different locations of a single unit in one of these ``rows$"$. Written down  in this basis explicitly, the multiphoton wave function can be used for constructing the density matrix $\rho^{(n)}(\{\sigma_i\},\{\sigma_i^\prime\})=\Psi^{(n)}(\{\sigma_i\})\Psi^{(n)\,\dag}(\{\sigma_j^\prime\})$.
However at high values of the photon numbers $n$ this procedure is rather cumbersome to be reproduced explicitly. Fortunately, there is a much more compact algorithm for constructing multiphoton density matrices to be described and discussed below. But of course, at any given $n$ correctness of the used below matrix representations can be checked and confirmed directly by the described derivations based on the use of the multiphoton wave functions $\Psi^{(n)}(\{\sigma_i\})$.

Thus, for any pure two-mode multiphoton state $\ket{\Psi^{(n)}}$ its $2^n\times 2^n$  density matrix can be presented symbolically in the following form
\begin{equation}
 \label{rho}
 \rho^{(n)}=\frac{1}{n!}
 \left(\Big\{\braket{(a_H^\dag)^{n-k_2}(a_V^\dag)^{k_2}a_H^{n-k_1}a_V^{k_1}}\Big\}\right)
\end{equation}
with averaging understood as $\braket{...}=\bra{\Psi^{(n)}}...\ket{\Psi^{(n)}}$. Such mean
products of the creation and annihilation operators can be referred to as correlators. The integers $k_1$ and $k_2$ (both $\geq 0$ and $\leq n$) in Eq. (\ref{rho}) numerate, correspondingly, groups of columns and rows in the matrix. At any given values of $k_1$ and $k_2$ columns and rows repeat themselves $C_n^{k_{1,2}}$ times, where $C_n^k=n!/k!(n-k)!$ are the binomial coefficients. Note also that the total powers of creation operators and total powers of annihilation operators in all elements are the same: $(n-k_2)+k_2=n$ and $(n-k_1)+k_1=n$. But proportions between powers of the creation operators in the $H$- and $V$-modes change from one line of the matrix to another and they are controlled by the integer $k_2$. Similarly, proportions between powers of the annihilation operators in the $H$- and $V$-modes change from one column of the matrix to another and they are controlled by the integer $k_1$.

The simplest examples are the density matrices of pure one-photon and two-photon polarization states
\begin{eqnarray}
\label{rho-one-ph}
 \nonumber
 k_1=\;\;\; 0\quad\quad\quad\; 1\;\;\quad\quad k_2\\
 \rho^{(1)}=\left(
 \begin{array}{cc}
 \braket{a_H^\dag  a_H} & \braket{a_H^\dag a_V}\\
 \braket{a_V^\dag a_H} &  \braket{a_V^\dag a_V}
\end{array}
 \right)
 \begin{array}{c} 0 \\ 1 \end{array}
\end{eqnarray}
and

\begin{eqnarray}
 \nonumber
  \rho^{(2)}={\small\frac{1}{2}}\times\quad\quad\quad\quad\quad\quad\quad\quad \\
\tiny{
 \nonumber
 k_1=\;\;\; 0\quad\quad\quad\quad\quad\quad 1\quad\quad\quad\quad\quad\quad\quad 1\quad\quad\quad\quad\quad\quad 2\quad\quad\quad\quad k_2}\\
 \times\tiny{\left(\begin{array}{cccc}
 \label{rho-biph}
  \braket{a_H^{\dag\,2} a_H^2}&\braket{a_H^{\dag\,2}a_H a_V}&\braket{a_H^{\dag\,2}a_H a_V}&\braket{a_H^{\dag\,2}a_V^2}\\
 \braket{a_H^\dag a_V^\dag a_H^2}&\braket{a_H^\dag a_V^\dag a_H a_V}&\braket{a_H^\dag a_V^\dag a_H a_V}&\braket{a_H^\dag a_V^\dag a_V^2}\\
 \braket{a_H^\dag a_V^\dag a_H^2}&\braket{a_H^\dag a_V^\dag a_H a_V}&\braket{a_H^\dag a_V^\dag a_H a_V}&\braket{a_H^\dag a_V^\dag a_V^2}\\
 \braket{a_V^{\dag\,2} a_H^2}&\braket{a_V^{\dag\,2} a_H a_V}&\braket{a_V^2 a_H a_V}&\braket{a_V^{\dag\,2} a_V^2}
 \end{array}
 \right)}
 \small{\begin{array}{c}
  0 \\ 1 \\ 1 \\ 2
 \end{array}
 }
\end{eqnarray}
and so on.

As mentioned above, the biphoton density matrix $\rho^{(2)}$ (\ref{rho-biph}) was written down by Klyshko \cite{Klyshko} and used in Refs. \cite{Klyshko, Che-Burl, Che-Kul}. Note however that the next step used for working with the density matrix (\ref{rho-biph}) consisted in simple crossing out one of two coinciding rows and one of two coinciding columns. This reduces the 4th-order matrix to the 3-dimensional one, but it changes significantly features of the arising matrix. In particular, its trace becomes different from one in contrast to the density matrix (\ref{rho-biph}). Also it does not provide a correct transition to the so called coherence matrix of biphoton qutrits \cite{JETP}. Indeed, the most general polarization biphoton state is qutrit, the sate vector of which is
\begin{equation}
 \label{qutrit}
 \ket{\Psi^{(2)}}=C_1\ket{2_H}+C_2\ket{1_H,1_V}+C_3\ket{2_V}
\end{equation}
with $C_i$ being arbitrary complex constants obeying the normalization condition $\sum_i|C_i|^2=1.$ The natural coherence matrix of this state is
\begin{eqnarray}
\rho_{coh}^{(2)}=
 \nonumber
 {\small
 \left(\begin{array}{ccc}
 |C_1|^2&C_1^*C_2&C_1^*C_3\\
 C_2^*C_1&|C_2|^2&C_2^*C_3\\
 C_3^*C_1&C_3^*C_2&|C_3|^2
 \end{array}
  \right)}=\quad\\
  \label{coherence}
 {\small
 \left(\begin{array}{ccc}
 \frac{\braket{a_H^{\dag\,2} a_H^2}}{2}&\frac{\braket{a_H^{\dag\,2}a_H a_V}}{\sqrt{2}}&\frac{\braket{a_H^{\dag\,2}a_V^2}}{2}\\
 \frac{\braket{a_H^\dag a_V^\dag a_H^2}}{\sqrt{2}}& \braket{a_H^\dag a_V^\dag a_H a_V}&\frac{\braket{a_H^\dag a_V^\dag a_V^2}}{\sqrt{2}}\\
  \frac{\braket{a_V^{\dag\,2} a_H^2}}{2}& \frac{\braket{a_V^{\dag\,2}a_H a_V}}{\sqrt{2}}&\frac{\braket{a_V^{\dag\,2}a_V^2}}{2}
 \end{array}
  \right)}.
 \end{eqnarray}
Evidently, the last expression (\ref{coherence}) does not coincide with that of Eq. (\ref{rho-biph}) with, e.g., deleted the third column and third row. So, the procedure of crossing out repeated columns and rows can not be considered as mathematically correct. To make it correct, one has to make first the unitary transformation of the matrix (\ref{rho-biph}) \cite{JETP}, after which all elements in one of rows and one of columns in the $4\times 4$ matrix turn zero. For the matrix (\ref{rho-biph}) the required unitary transformation has the form

$$
\rho^{(2)}\rightarrow {\widetilde\rho}^{(2)}=U\rho^{(2)}U^\dag
$$
with
$$
U=
{\small
%\begin{eqnarray}
\left(\begin{array}{cccc}
1&0&0&0\\
0&1/\sqrt{2}&1/\sqrt{2}&0\\
0&1/\sqrt{2}&-1/\sqrt{2}&0\\
0&0&0&1\end{array}\right)
%\end{eqnarray}
}.
$$

Only after this transformation the arising single line and single column with zero elements can be safely removed without changing general features of the original density matrix and providing the correct expression for the coherence matrix (\ref{coherence}) \cite{JETP}. In principle, similar transformations can be found also for density matrices of higher-order states, $n>2$. But in the following discussion we will not use such transformations by keeping all the full $2^n$ dimensionality of the density matrices unchanged, with repeating identical columns and rows of the density matrix completely conserved. Actually this repetition of columns and rows is related directly with the symmetry features of multi-boson wave functions. This symmetry is not seen explicitly in the multiphoton state vectors of the type (\ref{basis}), but in the wave function of polarization variables they are present in the form of terms differing only by transposition of variables \cite{Schweber,JETP,NJP}. Such terms in the wave functions are responsible directly for appearance of repeated columns and rows in the density matrices.

\section{Reduced density matrices}
As known the degree of entanglement of pure quantum states is related directly to the degree of mixing of reduced state. The concept of reduced states arises when one represents a complicated pure states as if consisting of two parts. And then reduction is averaging over one of these two parts giving rise to possibly mixed state of the other part. In the simplest cases of $n=2$ and $n=3$ definitions of two parts are evident: these parts consist of two single-photon states in the case of biphotons, and they consist of a single-photon and two-photon states in the case of a pure three-photon original states. In the cases of states with large numbers of photons, $n\geq 4$,  there are more than one ways of imagining how the original $n$-photon state can be divided for two parts. E.g. for $n=4$ there are are two ways of the gedanken splitting this state for two parts: $4=2+2$ and $4=3+1$ \cite{Mikl}. Thus, in these cases one can speak about different degrees of entanglement corresponding to different ways of splitting the original state for two parts.

Mathematically, a standard way of reducing density matrices of pure states consists in using their wave-function representation $\rho^{(n)}(\{\sigma_i\},\{\sigma_i'\})=\Psi^{(n)}(\{\sigma_i\})\Psi^{{(n)}\,\dag}(\{\sigma_i'\})$, equalizing one or several variables $\sigma_i=\sigma_i'$ and summing the product $\Psi^{(n)}\Psi^{{(n)}^\dag}$ over the variable(s) $\sigma_m$. But the procedure is rather cumbersome for states with many photons and with all symmetry requirements to the multi-boson wave functions completely taken into account. Fortunately, the result of such calculations can be presented in a relatively simple form with elements of the reduced density matrices expressed in terms of correlators similar to those arising in the described above density matrices of pure states. By assuming that for an $n$-photon state we reduce the density matrix $\rho^{(n)}$ with respect to $n-m$ variables, we can write the following general expression for the resulting reduced $2^m$-order density matrix
\begin{equation}
 \label{rho-red-m}
 \rho_r^{(m;\,n)}=\frac{(n-m)!}{n!}
 \left(
 \Big\{\braket{(a_H^\dag)^{m-k_2}(a_V^\dag)^{k_2}a_H^{m-k_1}a_V^{k_1}}\Big\}
 \right)
\end{equation}
with the previous definition of averaging in correlators $\braket{...}=\bra{\Psi^{(n)}}...\ket{\Psi^{(n)}}$ and with the previous meaning of the integers $k_2$ and $k_1$ ($m\geq k_{1,2}\geq 0$) numerating groups of columns and rows, at given $k_1$ and $k_2$ repeated $C_m^{k_{1,2}}$ times. Below are some examples of the reduced matrices.

The single-photon reduced density matrices of arbitrary pure $n$-photon states $\ket{\Psi^{(n)}}$ arising at $m=1$ have the form
\begin{equation}
 \label{rho-r-n-1}
 \rho_r^{(1;\,n)}=
 \frac{1}{n}\left(\begin{array}{cc}
 \braket{a_H^\dag a_H}&\braket{a_H^\dag a_V}\\
 \braket{a_V^\dag a_H}&\braket{a_V^\dag a_V}
 \end{array}
 \right).
\end{equation}
For basic Fock states $\ket{\Psi_{n_H,\,n_V}}$ (\ref{basis}) these matrices are very simple
\begin{equation}
 \label{rho-r-n-1-basis}
 \rho_r^{(1;\,n)}=\frac{1}{n_H+n_V}
 \left( \begin{array}{cc}
 n_H & 0\\
 0 & n_V
 \end{array}\right),
\end{equation}
and they correspond to the Schmidt entanglement parameter
\begin{equation}
 \label{k 1+n-1}
 K(n_H,\,n_V)=\frac{1}{Tr[(\rho_r^{(1;\,n)})^2]}=\frac{(n_H+n_V)^2}{n_H^2+n_V^2}.
\end{equation}
In the case of even total numbers of photon $n=n_H+n_V$, as a function of $n_H$, the Schmidt parameter $K$ achieves maximum at $n_H=n_V=n/2$ and $K_{\rm max}$ =2. At other relations between of $n_H$ and $n_V$ the Schmidt parameter $K$ is smaller than $K_{\rm max}$. In the cases of odd numbers of photons $n$ the maximal values of the Schmidt parameter are achieved at $n_H=[n/2]$ and $n_H=[n/2]+1$, where the symbol $[x]$ denotes in this case the integer closest to but smaller than $x$. Maximal values of the Schmidt parameter in these cases are somewhat smaller than 2. The simplest example of the basic Fock state with odd $n$ is that of three-photon states $\ket{1_H,2_V}$ and $\ket{2_H,1_V}$. In both cases Equation (\ref{k 1+n-1}) gives $K=9/5$ in agreement with the results of the work \cite{Mikl}. The main conclusion from this brief analysis concerns achievable entanglement of $n$-photon basic Fock states with respect to division for subsystems of a single-photon and an $(n-1)$-photon states: entanglement of such states with respect to such division for subsystems does not exceed that occurring in the case of biphoton states, and the maximal entanglement with $K=2$ or close to 2 is achieved in the states with maximally close numbers of horizontally and vertically polarized photons, $n_H$ and $n_V$.

The two-photon reduced density matrices of arbitrary pure $n$-photon states $\Psi^{(n)}$ arise in the cases of $m=2$ and their general form is given by
\begin{eqnarray}
 \nonumber
 \rho_r^{(2;\,n)}=\frac{1}{n(n-1)}\times\quad\quad\quad\quad\quad\\
 \label{rho-biph-r-n}
 \tiny{\left(\begin{array}{cccc}
  \braket{a_H^{\dag\,2} a_H^2} & \braket{a_H^{\dag\,2}a_H a_V} & \braket{a_H^{\dag\,2}a_H a_V} & \braket{a_H^{\dag\,2}a_V^2}\\
 \braket{a_H^\dag a_V^\dag a_H^2} & \braket{a_H^\dag a_V^\dag a_H a_V} & \braket{a_H^\dag a_V^\dag a_H a_V} & \braket{a_H^\dag a_V^\dag a_V^2}\\
 \braket{a_H^\dag a_V^\dag a_H^2} & \braket{a_H^\dag a_V^\dag a_H a_V} & \braket{a_H^\dag a_V^\dag a_H a_V} & \braket{a_H^\dag a_V^\dag a_V^2}\\
 \braket{a_V^{\dag\,2} a_H^2} & \braket{a_V^{\dag\,2} a_H a_V} & \braket{a_V^{\dag\,2} a_H a_V} & \braket{a_V^{\dag\,2} a_V^2}
 \end{array}\right)}.
\end{eqnarray}
Formally, this density matrix looks identical to that of Equation (\ref{rho-biph}), though normalization factors in these two matrices are different. But even more important difference concerns the meaning of averaging in correlators in these matrices. If the case of the density matrix of a pure two-photon states $\rho^{(2)}$ (\ref{rho-biph}) averaging is defined as $\braket{\Psi^{(2)}|...|\Psi^{(2)}}$. In contrast, in the case of the second-order reduced density matrix (\ref{rho-biph-r-n}) correlators in this matrix are defined as $\braket{\Psi^{(n)}|...|\Psi^{(n)}}$, where $n>2$. Note also that all described matrices, both of pure states (\ref{rho})-(\ref{rho-biph}) and of mixed states (\ref{rho-red-m})-(\ref{rho-biph-r-n}), obey the same important feature: their traces are equal to one.

For evaluating the degree of entanglement of multiphoton states $\ket{\Psi^{(n)}}$ their reduced density matrices have to be diagonalized numerically after which the found eigenvalues $\lambda_i^{(m;\,n)}$ can be used for finding the Schmidt entanglement parameter or the entropy of the reduced density matrices
\begin{equation}
 \label{K-S}
 K=\frac{1}{\sum_i\lambda_i^2}\quad {\rm and}\quad
 S_r=-\sum_i\lambda_i \log_2\lambda_i.
\end{equation}

Before presenting specific results of calculations, it's worth making a note concerning features of the described above density matrices and differences between their features in the cases of basic Fock states (\ref{basis}) and their superpositions (\ref{Sup}). In the case of single basic Fock states their pure-state and reduced density matrices have many zeros. In fact, averaging over basic Fock zeroes all correlators containing products of creation and annihilation operators in one of two modes in different powers, e.g., such as  $(a_H^\dag)^p\,a_H^q$ with $p\neq q$ and the same for the vertical-polarization mode. Owing to this, the density matrices of single Fock states turn out  having a diagonal-block structure. The following Equation represents an example of such a diagonal-block second-order reduced density matrix $\rho_r^{(2;\,4)}$ (\ref{rho-biph-r-n}) for the state $\ket{2_H,2_V}$ reduced with respect to two variables ($m=2$):
\begin{equation}
 \label{diag-bl-4-2}
 \rho_r^{(2;\,4)}=\left(
 \begin{array}{cccc}
 1/6&0&0&0\\
 0&1/3&1/3&0\\
 0&1/3&1/3&0\\
 0&0&0&1/6
 \end{array}
 \right)
\end{equation}
In this matrix three diagonal blocks are located (a) at the crossing of the first line and first column, (b) at the crossing of the 2nd and 3rd lines with the 2nd and 3rd columns and (c) at the crossing of the 4th line and 4th column. Each block gives only one non-zero eigenvalue, and they are  equal to, correspondingly, 1/6, 2/3, and 1/6, which gives $K=2$ in accordance with the result shown in Figure \ref{fig1}.

In a general case of the reduced density matrices $\rho_r^{(m;\,n)}$ (\ref{rho-red-m}) corresponding to the original states $\Psi_{n_H,n_V}$ (\ref{basis}) the non-zero square blocks  arise at crossings of the lines and columns with equal numbers of integers $k_1$ and $k_2$, $k_1=k_2\equiv k$ with $0\leq k\leq m$, and the dimensionality of each such blocks is $C_m^k$. The number of bloks equals to $m+1$.  All elements inside each block are equal to each other. Owing to equality of elements inside a block, each block has only one nonzero eigenvalue, and eigenvalues of the reduced density matrix can be expressed via these non-zero eigenvalues of blocks. Explicitly they are given by
\begin{eqnarray}
\nonumber
\lambda_k=\frac{(n-m)!}{n!}C_m^k\braket{\Psi_{n_H,n_V}|(a_H^\dag a_H)^{m-k}(a_V^\dag a_V)^k|\Psi_{n_H,n_V}}\\
 \label{eigenval}
 =\frac{(n-m)!}{n!}\frac{m!}{k!(m-k)!}\frac{n_H!}{(n_H-m+k)!}\frac{n_V!}{(n_V-k)!},\quad\quad\quad
\end{eqnarray}
with additional limitations
\begin{eqnarray}
 \nonumber
 k\leq\min\{ n_V,m\}\quad\quad {\rm and}\quad\quad\quad\quad\\
 \label{limitations}
 k\geq \max\{m-n_H,0\}\equiv\max\{n_V-(n-m),0\}.
\end{eqnarray}
Notice that at $m=n=n_H+n_V$ the reduced matrix $\rho_r^{(m;n)}$ turns into the density matrix of a pure state $\rho^{(n)}$. In this case the limitations  (\ref{limitations}) take the form $k\leq n_V$ and $k\geq n_V$, and they are compatible with each other only at $k=n_V$. This means that at a given value of $n_V$ the density matrix $\rho^{(n)}$ has only one nonzero block characterized by $k=n_V$. A simple algebra shows that in this case Equation (\ref{eigenval}) yields $\lambda_k=1$ as it has to be for a pure state.

The described features of the reduced density matrices corresponding to the basic two-mode Fock states $\Psi_{n_H,n_V}$ (\ref{basis}) simplify significantly diagonalization of these matrices and their Schmidt-mode analysis. The situation appears to be absolutely different in the case of superpositions of basic states $\Psi^{(n)}$ (\ref{Sup}). In this case the diagonal-block structure of matrices does not exist anymore and the reduced density matrices have to be diagonalized without any helping simplifications.

\section{Results}

\begin{figure}[h]
  \centering
  \includegraphics{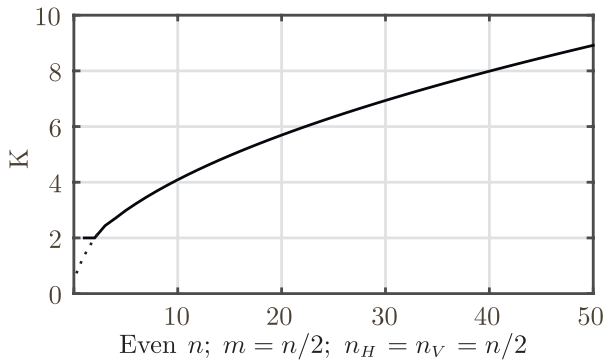}
  \caption{The calculated Schmidt entanglement parameter $K(n)$ for states $\ket{\Psi^{(n)}}=|n_H,n_V\rangle$ with even $n$, equal numbers of horizontally and vertically polarized photons, $n_H=n_V=n/2$, and with the gedanken splitting of the states for two $m$-photon states with equal numbers of photons, $m=(n-m)=n/2$; the dotted line corresponds to $K_{\rm appr}$ of Equation (\ref{approx})}\label{fig1}
\end{figure}

\begin{figure}[b]
  \centering
  \includegraphics{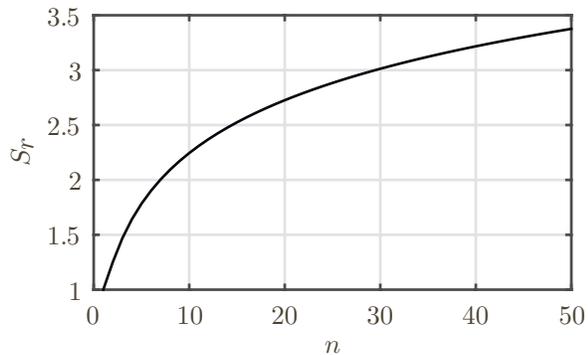}
  \caption{The same as in Figure \ref{fig1} but for the entropy of the reduced state rather than for the Schmidt entanglement parameter}\label{fig2}
\end{figure}

\begin{figure}[b]
  \centering
  \includegraphics{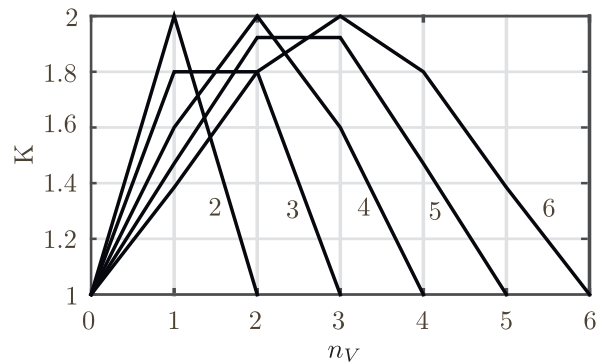}
  \caption{The Schmidt entanglement parameter $K$ as a function of the number $n_V$ of vertically polarized photons in the states $\ket{\Psi^{(n)}}$; total numbers of photons $n$ are shown near the curves; the assumed division for subsystems is $n\rightarrow 1+(n-1)$}\label{fig3}
\end{figure}

\begin{figure}[h]
  \centering
  \includegraphics{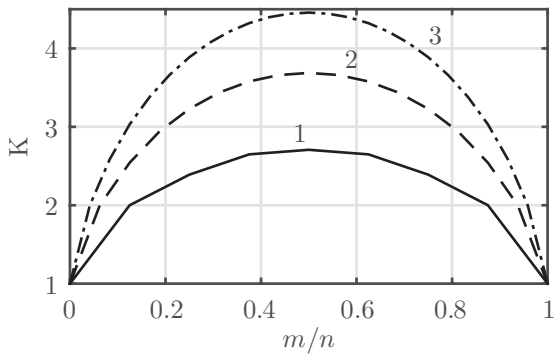}
  \caption{The Schmidt entanglement parameter $K$ of the states $\ket{\Psi_{n_H,n_V}}$ (\ref{basis}) vs. the ratio ``the number $m$ of variables in the reduced density matrix divided by the total number of polarization variables or the total number of photons $n"$; the curves correspond to $n=6\,(1),\,8\,(2)\,{\rm and}\,24\,(3)$}\label{fig4}
\end{figure}

\begin{figure}[h]
  \centering
  \includegraphics{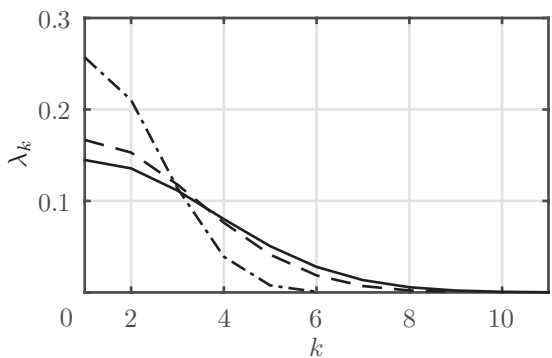}
  \caption{Arranged in descending order, eigenvalues $\lambda_k$ of the reduced density matrices $\rho^{(m,n)}_r$ (\ref{rho-red-m}) of the state $\ket{\Psi_{n_H,n_V}}$ (\ref{basis}) with $n_H=n_V=60$ and $n=n_H+n_V=120$ and different degrees of reduction: $m=50$ (solid line), 30 (dashed line) and 10 (dash-dotted line)}\label{fig5}

\end{figure}

\begin{figure}[h]
  \centering
  \includegraphics{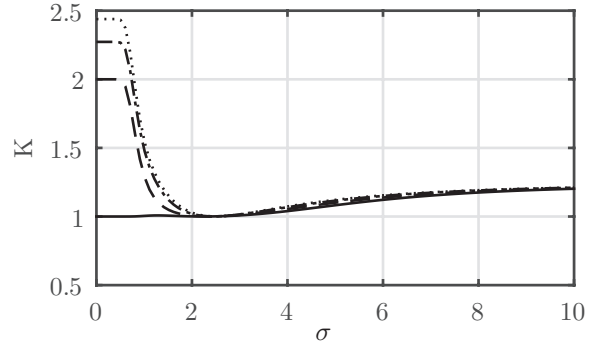}
  \caption{The Schmidt entanglement parameter $K$ for the state (\ref{superpos}) with $n=6$ and $m_0=3,\,2,\,1,\,0$ (from top to down at small values of $\sigma$)}\label{fig6}
\end{figure}

The results of calculations are presented in a series of pictures of Figures  \ref{fig1}-\ref{fig6}. The first of these pictures (Figure \ref{fig1})
corresponds to multiphoton states $\ket{\Psi^{(n)}}$ with the total number of photons $n$, where $n$ is taken even, and with equal numbers of photons with horizontal and vertical  polarizations, $n_H=n_V=n/2$. The state is assumed to be imagined consisting of two parts with the same numbers of photons in each, $n/2$. The reduced density matrix of such subsystem is $\rho^{(\frac{n}{2},\,n)}$ ($m=n-m=n/2$ in notations of Equation (\ref{rho-red-m})). Its eigenvalues are $\lambda_i$ and the Schmidt entanglement parameter is determined by the first expression in Equation (\ref{K-S}). In Figure \ref{fig1} the Schmidt parameter is shown in its dependence on the total number of photons in the state $\ket{\Psi^{(n)}}$. As seen from the picture of Figure \ref{fig1}, in the considered case the Schmidt entanglement parameter and, hence, the degree of entanglement are monotononically growing function of the number of photons. In other words, multiphoton Fock states can have much higher resource of entanglement than usually considered biphoton states.

Note  that the curve in Figure \ref{fig1} can be perfectly approximated by the analytical expression
\begin{equation}
 \label{approx}
 K_{\rm appr}\approx 0.62+n^{0.54}.
\end{equation}
Coincidence of this model curve with the numerically calculated one is so perfect that in the picture of Figure \ref{fig1} they look indistinguishable, except for a small region $n<4$. The main qualitative conclusion from this comparison is that as a function of the total number of photons $n$, the Schmidt entanglement parameter $K(n)$ grows roughly as the root square of $n$.

Similar conclusions can be deduced from calculations of the entropy of reduced state $S_r$ defined by the second expression in Equations (\ref{K-S}). For the same state as in the previous calculations the function $S_r(n)$ plotted in Figure \ref{fig2} is seen to be monotonically growing and being very similar to the curve of Figure \ref{fig1}. This confirms the conclusion about growing degree of entanglement with the growing number of photons and confirms compatibility of the entropy and Schmidt parameter for characterization of the degree of entanglement.

The picture of Figure \ref{fig3} describes dependencies of the Schmidt entanglement parameter $K$ on the relation between horizontally and vertically polarized photons in the Fock states with given total numbers of photons ${n}$: if the number of vertically polarized photons is $n_V=k\leq n$, the number of horizontally polarized photons is $n_H=n-k$, and the number $k$ varies along the horizontal axis in the picture of Figure \ref{fig3}. In this series of calculations the degree of reduction is taken to be as high as possible, $m=1$, i.e., the reduced state is a single-photon one and its reduced density matrix is $\rho_r^{(1;\,n)}$ of Equation (\ref{rho-r-n-1}).
As seen well from the pictures at all values of $n$ the Schmidt number $K$ and the degree of entanglement are maximal when the numbers of vertically and horizontally polarized photons in the state $\ket{\Psi^{(n)}}$ are equal ($k=n/2$) or maximally close to each other (in the case of odd $n$).

The picture of Figure \ref{fig4} shows the dependence of the Schmidt entanglement parameter of the Fock state $\ket{\Psi^{(n)}}$ on $m/n$, i.e., on the ratio of number $m$ of variables remaining in the state after its reduction to the total number of photons (or their variables) $n$ in the original pure state.
The picture shows clearly that entanglement of the state $\ket{\Psi^{(n)}}$ is maximal when it is considered as split for two parts with equal number of photons in each parts ($m/n=0.5$).

The picture of Figure \ref{fig5} shows the dependence of eigenvalues $\lambda_k$ on their numbers $k$ for the reduced density matrices $\rho_r^{(m;\,n)}$ of the state with the total number of photons $n=120$, $n_H=n_V$ and different degrees of reduction $n-m$.

The results shown in Figure \ref{fig5} show that in spite of a growing degree of entanglement in strongly multiphoton states, eigenvalues of all reduced density matrices remain concentrated in a restricted region of not too high values. This means that the effective dimensionality of the corresponding Hilbert spaces remains not too high. This conclusion is important for approximate numerical calculations because it opens a possibility of performing these calculations in smaller-  dimensionality matrices forming the main cores for finding relatively large eigenvalues $\lambda_k$.

Let us consider now an example of states more complicated than a single basic Fock state. Let the state under consideration be given by
\begin{equation}
 \label{superpos}
 \ket{\Psi}=\sum_{m=1}^n C_m\ket{(n-m)_H,m_V}.
\end{equation}
Let us take the coefficients $C_m$ in the Gaussian form
\begin{equation}
 \label{Gauss}
 C_m=N\exp{\left(-\frac{(m - m_0)^2}{2\sigma^2}\right)}.
\end{equation}
with the  normalization factor $N$ given by
\begin{equation}
 \label{norm}
 N=\left[\sum_{m=0}^{n}\exp{\left(-\frac{(m - m_0)^2}{\sigma^2}\right)}\right]^{-1/2}
\end{equation}
and $m_0$ is that value of $m$  at which the squared coefficients $|C_m|^2$ are maximal.
As mentioned above in this case diagonalization of the reduced density matrix is more complicated because this matrix does not have anymore a diagonal-block structure, and it has to be diagonalized as a whole, without any simplifications. Nevertheless, the results of such calculations are presented in Figure  \ref{fig6} for three different values of the parameter $m_0$ in the Gausssian distribution of Equation (\ref{Gauss}).

One of the most interesting features of the curves in Figure \ref{fig6} concerns disappearance of entanglement ($K=1$) at some definite point $\sigma_0$. In principle, this does not contradict, e.g., to the known features of the simplest superposition of Fock states - biphoton polarization qutrit (\ref{qutrit}) characterized by three constants $C_1$, $C_2$, $C_3$. As known \cite{NJP}, its degree of entanglement can be characterized either by the Schmidt entanglement parameter $K$ or by the so called concurrence $C=|2C_1C_3-C_2^2|$ \cite{Wootters}, which are related to each other by a simple formula $C=\sqrt{2(1-K^{-1})}$. It's known also that entanglement of qutrit disappears when $C=0$ or $2C_1C_3=C_2^2$. This effect of  disappearing entanglement at some specific relation between the qutrit's parameter seems to be analogous to the effect of missing entanglement of the state (\ref{superpos}) at $\sigma = \sigma_0$

\section{Conclusion}

Thus, in this paper the density-matrix approach used earlier for biphoton states is generalized for the case of multiphoton two-mode polarization states. Both pure two-mode Fock states and their superpositions with given total numbers of photons are considered.  In this method elements of density matrices are expressed in terms of mean values of products  of photon creation and annihilation operators.Structures of the arising density matrices reduced with a part of polarization variables is discussed. Eigenvalues $\lambda_k$ of the reduced density matrices are found analytically for Fock states and numerically for their superpositions.These results are used for finding the degree of entanglement of multiphoton states with respect to their division for pairs of states with smaller numbers of photons. The degree of entanglement is estimated either by the Schmidt entanglement parameter $K=1/\sum_k\lambda_k^2$ or by the entropy of the reduced states $S=-\sum_k\lambda_k\log_2\lambda_k$. The main qualitative conclusion is that the degree of entanglement is maximal if numbers of photon in two modes, $n_H$ and $n_V$, are maximally close to each other and if multiphoton states are considered as consisting to two parts with approximately (or exactly) equal numbers of photons in each of two parts. The maximal degree of entanglement is found to be a growing function of the number of photons as shown in Figures \ref{fig1} and \ref{fig2}.

\section*{Acknowledgement}
The work is supported by the Russian Foundation for Basic Research, grant 18-02-00634.

\bibliography{text}

%\pagebreak

\end{document}